\documentclass{elsart}

\usepackage{graphicx}
\usepackage{amssymb}

\begin{document}

\begin{frontmatter}
\title{The effects of nonlinear couplings and  external magnetic field on the thermal entanglement in a two-spin-qutrit
system}
\author[label1]{Guo-feng Zhang \corauthref{cor1}},
\corauth[cor1]{Corresponding author.}
\ead{gf1978zhang2001@yahoo.com}
\author[label2]{Shu-shen Li}
\address[label1]{State Key Laboratory for Superlattices and
Microstructures, \\ Institute of Semiconductors, Chinese Academy
of Sciences, P. O. Box 912, Beijing 100083, P. R. China}
\address[label2]{China Center of Advanced Science and Technology
(CCAST) (World Laboratory), P.O. Box 8730, Beijing 100080, China
and State Key Laboratory for Superlattices  and Microstructures,
Institute of Semiconductors, Chinese Academy of Sciences, P.O. Box
912, Beijing 100083,China}

\begin{abstract}
\quad We investigate the effects of nonlinear couplings and
external magnetic field on the thermal entanglement in a
two-spin-qutrit system by applying the concept of negativity. It
is found that the nonlinear couplings favor the thermal
entanglement creating. Only when the nonlinear couplings $|K|$ are
larger than a certain critical value does the entanglement exist.
The dependence of the thermal entanglement in this system on the
magnetic field and temperature is also presented. The critical
magnetic field increases with the increasing nonlinear couplings
constant $|K|$. And for a fixed nonlinear couplings constant, the
critical temperature is independent of the magnetic field $B$.
\end{abstract}

\end{frontmatter}

% main text
\section{Introduction}
\quad Entanglement, first noted by Schr\"{o}dinger \cite{sch} and
Einstein, Podolsky, and Rosen \cite{epr}, is an essential feature
of quantum mechanics. In the last decades it was rediscovered as a
new physical resource for quantum information processing (QIP).
Since the entanglement is very fragile, the problem of how to
create stable entanglement remains a main focus of recent studies
in the field of QIP. The thermal entanglement, which differs from
the other kinds of entanglement by its advantages of stability and
requires neither measurement nor controlled switching of
interactions in the preparing process, is an attractive topic that
has been extensively studied for various systems including
isotropic \cite{mca,kmo,xwa,gfz} and anisotropic \cite{xwan,lss}
Heisenberg chains, Ising model in an arbitrarily directed magnetic
field \cite{dgu}, and cavity-QED \cite{sma} since the seminal
works by Arnesen et al.\cite{mca} and Nielsen \cite{mani}. Based
on the method developed in the context of quantum information, the
relaxation of a quantum system towards the thermal equilibrium is
investigated \cite{vsc} and provides us an alternative mechanism
to model the spin systems of the spin-$\frac 12$ case for the
approaching of the thermal entangled states
\cite{mca,kmo,xwa,gfz,xwan}. But only spin-$\frac{1}{2}$ cases are
considered in the above papers. Zhang et al.\cite{gzh,gzh1}
investigated the thermal entanglement in the two-spin-$1$ system
with a magnetic field and gave a comparison between the uniform
magnetic field case and the nonuniform one. There the nonlinear
couplings is ignored for simplicity. In this paper, we will
investigate the effects of nonlinear couplings and external
magnetic field on the thermal entanglement in a two-spin-qutrit
system. Thus we may better understand and make use of entanglement
in QIP through changing the environment parameters.

\section{The model Hamiltonian and the solutions}
\quad The development of laser cooling and trapping provides us
more ways to control the atoms in traps. Indeed, we can manipulate
the atom-atom coupling constants and the atom number in each
lattice well with a very good accuracy. Our system consists of two
wells in the optical lattice with one spin-1 atom in each well.
The lattice may be formed by three orthogonal laser beam, and we
may use an effective Hamiltonian of the Bose-Hubbard form
\cite{dja}to describe the system. The atoms in the Mott regime
make sure that each well contains only one atom. For finite but
small hopping term $t$, we can expand the Hamiltonian into powers
of $t$ and get\cite{sky},
\begin{equation}
H=\epsilon +J(S_1\cdot S_2)+K(S_1\cdot S_2)^2,
\end{equation}
where $J=-\frac{2t^2}{U_2}$,
$K=-\frac{2t^2}{3U_2}-\frac{4t^2}{U_0}$ with $t$ the hopping
matrix elements, and $\epsilon =J-K$. $U_s$($s=0,2$) represents
the Hubbard repulsion potential with total spin $s$, a potential $U_s$ with $%
s=1$ is not allowed due to the identity of the bosons with one
orbital state per well, since the term $\epsilon $ contains no
interaction, we can ignore it in the following discussions and it
would not change the thermal entanglement. In this paper we will
confine ourself to the case of $K<0$ and $J<0$ that is relevant to
the recent experiment conducted on $^{23}N_{a}$ atoms. Here we
mainly investigate the effects of nonlinear couplings on the
entanglement. So $|K|\gg |J|$ is assumed, the Hamiltonian Eq.(1)
becomes
\begin{equation}
H=K(S_1\cdot S_2)^{2},
\end{equation}
with an external magnetic field, our system is described by
\begin{equation}
H=K(S_1\cdot S_2)^{2}+B(S_{1z}+S_{2z}),
\end{equation}
where $S_\alpha $ ($ \alpha =x,y,z$) are the spin operators, $K$
is the nonlinear couplings constant and the magnetic field is
assumed to be along the $z$-axes. When the total spin for each
site $S_j=1$ $(j=1, 2)$, its components take the form:
\begin{equation}
S_{jx}=\frac 1{\sqrt{2}}\left(
\begin{array}{lll}
0 & 1 & 0 \\
1 & 0 & 1 \\
0 & 1 & 0
\end{array}
\right) ,S_{jy}=\frac 1{\sqrt{2}}\left(
\begin{array}{lll}
0 & -i & 0 \\
i & 0 & -i \\
0 & i & 0
\end{array}
\right) ,S_{jz}=\left(
\begin{array}{lll}
1 & 0 & 0 \\
0 & 0 & 0 \\
0 & 0 & -1
\end{array}
\right).
\end{equation}

\quad In order to proceed we first of all find the eigenvalues and
the corresponding eigenstates of the Hamiltonian which are seen to
be
\begin{eqnarray}
H\left| \Psi _1\right\rangle &=&(K+2B)|\Psi_{1}\rangle, H\left|
\Psi _2\right\rangle =(K-2B)|\Psi_{2}\rangle\nonumber,
\\H\left|\Psi_3\right\rangle&=&(\frac{K}{4}+B)|\Psi_{3}\rangle,
H\left|\Psi_4\right\rangle=(\frac{K}{4}+B)|\Psi_{4}\rangle\nonumber,
\\H\left| \Psi_5\right\rangle&=&(\frac{K}{4}-B)|\Psi_{5}\rangle,
H\left|\Psi_6\right\rangle=(\frac{K}{4}-B)|\Psi_{6}\rangle\nonumber,
\\H\left|\Psi_7\right\rangle &=&K|\Psi_{7}\rangle,
 H\left|\Psi
 _8^{\pm}\right\rangle=\frac{2\pm\sqrt{3}}{2}K|\Psi_{8}^{\pm}\rangle.
\end{eqnarray}
where
\begin{eqnarray}
|\Psi_{1}\rangle&=&|1,1\rangle\nonumber,
|\Psi_{2}\rangle=|-1,-1\rangle\nonumber,
|\Psi_{3}\rangle=|1,0\rangle\nonumber,
\\|\Psi_{4}\rangle&=&|0,1\rangle\nonumber,
|\Psi_{5}\rangle=|0,-1\rangle\nonumber,
|\Psi_{6}\rangle=|-1,0\rangle\nonumber,
\\|\Psi_{7}\rangle&=&\frac{1}{\sqrt{2}}(|-1,1\rangle-|1,-1\rangle)\nonumber,
\end{eqnarray}
\begin{equation}
|\Psi_{8}^{\pm}\rangle=\frac{1}{\sqrt{2+(\sqrt{3}\mp1)^{2}}}(|1,-1\rangle+(1\mp\sqrt{3})|0,0\rangle+|-1,1\rangle).
\end{equation}
here $\left| \alpha, \beta\right\rangle $ $(\alpha=1, 0, -1$ and
$\beta=1, 0, -1)$ are the eigenstates of $S_{1z}S_{2z}$. The
density operator at the thermal equilibrium $ \rho (T)=\exp
(-\beta H)/Z$, where $Z=Tr[\exp (-\beta H)]$ is the partition
function and $\beta =1/k_BT$ ($k_B$ is Boltzmann's constant being
set to be unit $k_B=1$ hereafter for the sake of simplicity and
$T$ is the temperature), can be expressed in terms of the
eigenstates and the corresponding eigenvalues as
\begin{equation}
\rho =Z^{-1}\sum_{l}\exp [-\beta E_l]\left| \Psi _l\right\rangle
\left\langle \Psi _l\right|,
\end{equation}
where $E_l$  is the eigenvalue of the corresponding eigenstate and
the partition function  $Z=\exp[-\frac{K}{T}](1+2\cosh
[\frac{2B}{T} ]+2\cosh
[\frac{\sqrt{3}K}{2T}]+4\exp[\frac{3K}{4T}]\cosh [\frac{B}{T}])$.

\section{The negativity of the system}
\quad Here we will give the entanglement of the system by applying
the concept of negativity \cite{ape} which can be computed
effectively for any mixed state of an arbitrary bipartite system.
The negativity vanishes (i. e. negativity is equal to zero) for
unentangled states. For our purpose to evaluate the negativity, in
the following, we need to have a partially transposed density
matrix $\rho ^{T_A}$ of original density matrix $\rho $ with
respect to the eigenbase of any one spin particle ( say particle
$A$) in our two-spin system which is found in the basis $\{\left|
\alpha, \beta\right\rangle, \alpha=1, 0, -1$ and $\beta=1, 0,
-1\}$ as $\rho ^{T_A}=$
\begin{equation}
\frac 1Z\left(
\begin{array}{lllllllll}
e^{-\frac{K+2B}{T}} & 0 & 0 & 0 & 0 & 0 & 0 & 0
&\frac{M_{-}}{6}e^{-\frac{K}{T}}
 \\
0 & e^{-\frac{K+4B}{4T}} & 0 & 0 & 0 & Q & 0 & 0 & 0 \\
0 & 0 & \frac{M_{+}}{6}e^{-\frac{K}{T}} & 0 & 0 & 0 & 0 & 0 & 0 \\
0 & 0 & 0 & e^{-\frac{K+4B}{4T}} & 0 & 0 & 0 & Q & 0 \\
0& 0 & 0 & 0 & P & 0 & 0 & 0 & 0 \\
0 & Q & 0 & 0 & 0 & e^{-\frac{K-4B}{4T}}& 0 & 0 & 0 \\
0 & 0 & 0 & 0 & 0 & 0 & \frac{M_{+}}{6}e^{-\frac{K}{T}} & 0 & 0 \\
0 & 0 & 0 &Q & 0 & 0 & 0 & e^{-\frac{K-4B}{4T}} & 0 \\
\frac{M_{-}}{6}e^{-\frac{K}{T}} & 0 & 0 & 0 & 0 & 0 & 0 & 0 &
e^{-\frac{K-2B}{T}}
\end{array}
\right).
\end{equation}
where
$M{\pm}=\pm3+3\cosh[\frac{\sqrt{3}K}{2T}]-\sqrt{3}\sinh[\frac{\sqrt{3}K}{2T}]$,
$Q=\frac{\exp[-\frac{m}{T}]}{2\sqrt{3}}(-1+\exp[\frac{\sqrt{3}K}{T}])$,
$P=\frac{1}{6}\exp[-\frac{m}{T}](3-\sqrt{3}+(3+\sqrt{3})\exp[\frac{\sqrt{3}K}{T}])$
and $m=\frac{2+\sqrt{3}}{2}K$. The negativity as the entanglement
measure\cite{gvi} is defined by
\begin{equation}
N(\rho )=\frac{\left| \left| \rho ^{T_A}\right| \right| -1}2.
\end{equation}
where $\left|\left|\rho ^{T_A}\right|\right|=\sqrt{tr[\rho
^{T_A}]^{+}\rho ^{T_A}}$ denotes the trace norm\cite{kzy} of the
density matrix $\rho ^{T_A}$. The negativity $N(\rho)$ is
equivalent to the absolute value of the sum of the negative
eigenvalues of $\rho ^{T_A}$. From the definition $N(\rho
)=-\sum_{i}\lambda_{i}$ ($\lambda_{i}$ is the negative
eigenvalue), one can see that the maximum value of the absolute
value of the sum of the negative eigenvalues of $\rho ^{T_A}$ may
be greater than $1$ as long as $tr(\rho ^{T_A})=1$. For different
dimensions of density matrix, the maximum value is different. For
example, the maximum value is $0.5$ for the two qubits with two
levels. In our case, the negativity is $0.5$ in the state
$|\Psi_{7}\rangle$, the negativity is about $0.972$ in the state
$|\Psi_{8}^{+}\rangle$ and $0.683$ in the state
$|\Psi_{8}^{-}\rangle$, so the negativity value of the statistical
mixture of these states is less than $1$. In our plots, the value
$0.972$ means maximal entanglement.
\begin{figure}[h]
\begin{center}
\includegraphics[width=7 cm]{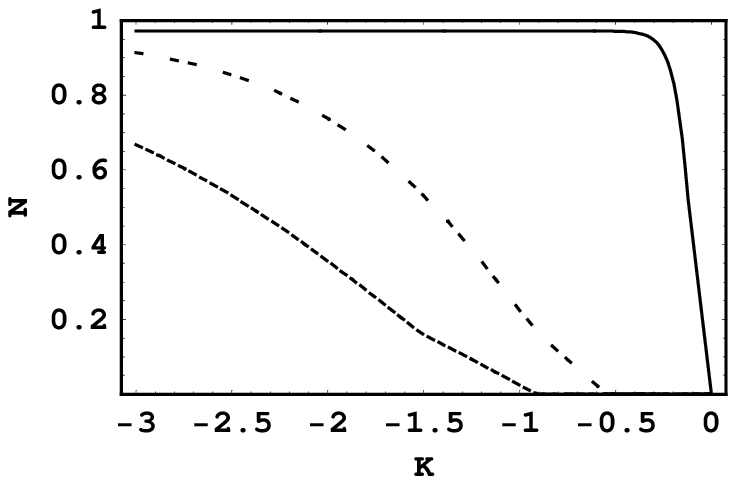}
\caption{The negativity vs. the nonlinear couplings constant $K$
for different temperature $T$ (from top to bottom, $T$ equals
$0.05$, $0.6$, $1.0$), the magnetic field $B=0$.}
\includegraphics[width=7 cm]{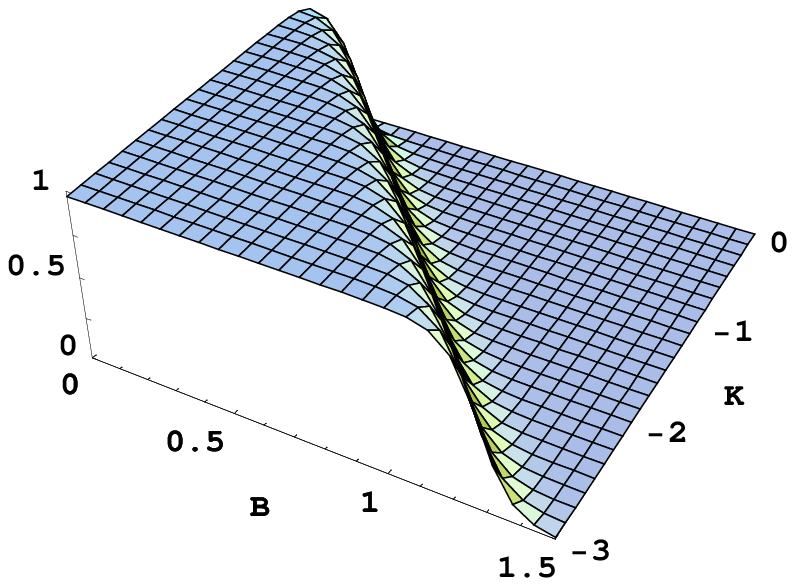}
\caption{(Color on line)The negativity vs. the nonlinear couplings
constant $K$ and the magnetic field $B$, the temperature $T=0.1$.}
\end{center}
\end{figure}

\quad We perform the numerical diagonalization of the density
matrix and the numerical results of the entanglement measure are
presented in figures from Fig.1-Fig.3. Fig.1 shows the plot of the
negativity as a function of the nonlinear couplings constant $K$
for different temperature $T$ when $B=0$. From the figure we can
see that the thermal entanglement increases with the increasing
$|K|$ and demonstrates a different evolvement curve for different
temperature. When $B=0$, $K<0$ and the temperature is close to
absolute zero, the state $|\Psi_{8}^{+}\rangle$ is seen to be the
ground state, which is a entangled state, so the negativity is not
equal to zero (in fact the negativity is equal to $0.972$) and it
decreases with the decreasing of $|K|$. Only when the nonlinear
couplings $|K|$ are larger than a certain critical value does the
entanglement exist. For $T=0.6$, the critical value of $|K|$ is
smaller than that for $T=1.0$. We can also see that the maximum
value that the negativity can arrive at gets smaller with the
increasing temperature for a fixed $|K|$ and the system can hold
high entanglement in a quite broader area of the nonlinear
couplings $K$ when the temperature is close to zero.

\quad The negativity as a function of the nonlinear couplings
constant $K$ and the magnetic field $B$ is plotted in Fig.2 when
$T=0.1$. The magnetic field plays a negative role for the
negativity at a fixed temperature, this can be seen from the
figure. There should exist a competition between the magnetic
field and the nonlinear couplings $|K|$. The critical magnetic
field for various $|K|$ is different and it increases nearly
linearly when $|K|$ increases. The negativity will arrive at zero
when the magnetic field has a very large value. This is very
easily understood since the term $K(S_1\cdot S_2)^{2}$ in the
Hamiltonian can be ignored when $B$ is very large, thus the system
is unentangled.
\begin{figure}
\begin{center}
\includegraphics[width=8 cm]{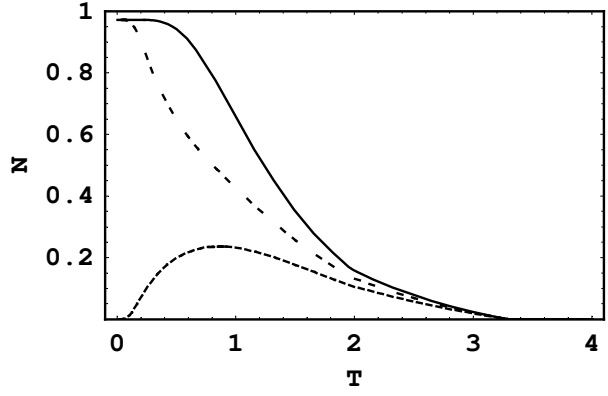}
\caption{The negativity vs. the temperature for different magnetic
field $B$ (from top to bottom, $B$ equals $0.2$, $1.0$, $1.5$),
nonlinear couplings constant $K=-3$.}
\end{center}
\end{figure}

\quad In Fig.3, the plot of the negativity as a function of the
temperature for different magnetic field is given. When $B$ is
close to zero, the  state $|\Psi_{8}^{+}\rangle$ is seen to be the
ground state in which the negativity is about $0.972$ at $T=0$. As
the temperature increases, $N$ rapidly decreases due to the mixing
of the excited states with the ground state. For a higher value of
the magnetic field $B$ (say $B=1.5$), the state $|\Psi_{2}\rangle$
becomes the ground state and there is no entanglement at $T=0$.
However we may increase the entanglement by increasing the
temperature $T$ in order to bring the entangled eigenstates such
as $|\Psi_{8}^{\pm}\rangle$ into mixing with the ground state. We
also found that the critical temperature is almost the same for
different external field. The change in negativity as $T$
increases from absolute zero is due to population of excited
levels. These results are consistent with those found in our
previous paper\cite{gfz,gzh}.

\section{Conclusions}
\quad We investigate qualitatively (not quantitatively) the
effects of nonlinear couplings and external magnetic field on the
thermal entanglement in a two-spin-qutrit system in terms of the
measure of entanglement called ``negativity". We find that the
nonlinear couplings favor the thermal entanglement. Only when the
nonlinear couplings $|K|$ are larger than a certain critical value
does the entanglement exist. Increasing the nonlinear couplings
$|K|$ increases the critical magnetic field, but the critical
temperature is almost the same for different external magnetic
field at a fixed nonlinear couplings constant.

\end{document}